# Microgrid Optimal Energy Scheduling with Risk Analysis

Ali Siddique
Department of Electrical and Computer Engineering
University of Houston
Houston, TX, USA
asiddique2@uh.edu

Cunzhi Zhao
Department of Electrical and Computer Engineering
University of Houston
Houston, TX, USA
czhao20@uh.edu

Xingpeng Li
Department of Electrical and Computer Engineering
University of Houston
Houston, TX, USA
xli82@uh.edu

*Abstract*—Risk analysis is currently not quantified in microgrid resource scheduling optimization. This paper conducts a conditional value at risk (cVaR) analysis on a grid-disconnected residential microgrid with distributed energy resources (DER). We assume the infrastructure to set up an ad-hoc microgrid is already in place for a residential neighborhood with power sources such as photovoltaic (PV), diesel, and battery energy storage system (BESS). With this scenario in mind, we solve day-ahead scheduling to optimally allocate various resources to match demand in scenarios where neighborhoods, especially residential, are disconnected from the overall grid such as in flooding, hurricanes, winter storms, or operational failures. The goal is to provide an alternative framework to optimize power availability for priority customers and strengthen the overall grid against dips in power outside of normal operating considerations. The focus of this paper will be taking in renewable energy sources from PV combined with diesel and BESS while minimizing cost. Case studies demonstrate that with the proposed energy management system, microgrids can be implemented to be more resilient against new challenges.

## Nomenclature

| Symbol | Description |
|---|---|
| $D_{p_t}$ | Priority customer demand defined as customers where electricity cannot be curtailed. |
| $D_{e_t}$ | Essential customer demand. Defined as residential customers whose electricity can be curtailed. |
| $D_{e_{c_t}}$ | Essential customer curtailed. Defined as residential customers whose electricity is curtailed. |
| $D_{Net\ Load_t}$ | Demand load of the system subtracted from any residential PV that is generated. |
| $D_{Load_t}$ | Demand total for all customers in microgrid. |
| $P_{BESS_t}$ | Power of the battery energy storage system. |
| $P_{PV_t}$ | Power value of photovoltaic residential solar panels. |
| $P_{Di_t}$ | Power output of diesel generator. |
| $P_{Di_{min}}$ | Power output minimum for diesel generator. |
| $P_{Di_{max}}$ | Power output maximum for diesel generator. |
| $P_{Total_t}$ | Total available power. |
| $P_{D_{max}}^B$ | The maximum discharge power of the battery. |
| $P_{D_t}^B$ | The discharging power of the battery. |
| $P_{C_t}^B$ | The charging power of the battery. |
| $P_{C_{max}}^B$ | The maximum charge power of the battery. |
| $C_{batt,red}$ | The additional price ($) of the battery cost when the battery state of charge is outside the green zone. |
| $C_{fuel}$ | Fuel cost ($/kW) of diesel generation. |
| $C_{B_{N_t}}$ | Degradation cost ($) of the battery. |
| $C_{B_{Total}}$ | The capital cost ($) of the battery. |
| $C_{D_e}$ | Cost ($/kW) of curtailing essential customers. |
| $U_r$ | Binary value indicating whether the battery state of charge is outside the green zone. |
| $U_{C_t}$ | Binary value for charging status. |
| $U_{D_t}$ | Binary value for discharging status. |
| $U_{di_t}$ | Binary state determining if diesel generator is on or off. |
| $SOC_t$ | The energy state of the battery. |
| $SOC_{min}^{green}$ | Minimum state of charge at which normal battery degradation can occurs. |
| $SOC_{max}^{green}$ | Maximum state of charge at which normal battery degradation can occurs. |
| $SOC_{min}$ | Minimum possible state of charge for battery. |
| $SOC_{max}$ | Maximum possible state of charge for battery. |
| $DoD_t$ | Depth of discharge for battery. |
| $DoC_t$ | Depth of charge for battery. |
| $N$ | Number of scenarios. |
| $N_{bat,max_t}$ | Rated maximum number of cycles of battery system. |
| $N_{bat_t}$ | Battery cycle count. |
| $cVaR$ | Conditional value at risk formulation used to calculate risk at high-risk low probability scenarios. |
| $VaR$ | Possible value at risk. |
| $\alpha$ | Smallest possible cost for admissible loss. |
| $f(x,y)$ | Unmet demand after generation is accounted. |
| $\beta$ | Confidence level. |
| $z_i$ | BESS power and diesel power of the specific interval. |
| $t$ | Time segment per analysis. |
| $\Delta\lambda$ | Change in degradation between two-time intervals. |
| $\lambda_{N_{bat_t}}$ | The capacity factor loss at the Nth cycle. |
| $\Delta T$ | The length of time segment. |
| $\gamma$ | Price normalizer value ($/cycle). |
| $\varepsilon$ | Relationship between essential and priority customers. |

## I. Introduction

There have been 500 weather events in North America impacting 50,000 customers for each event from 2005-2015 [1]. Similar increased electricity outages due to weather have been reported on other continents. These increases in the severity of natural disasters are due to the forces of climate change [2]. Also, blackouts have occurred due to operational errors resulting in millions of customers losing power [3]. Lastly, attacks against the grid have become more common from foreign actors [4]. Both trends have emphasized the need for a more distributed and decentralized electric grid which should function to some extent even if disconnected from the overall electric utility.

A microgrid is defined by the Department of Energy as "*a group of interconnected loads and distributed energy resources within clearly defined electrical boundaries that acts as a single controllable entity with respect to the grid*" [5].

Microgrid technology has become increasingly more common in the past few decades due to its ability to supply areas with geographical constraints, disaster prone issues, and rural areas. It is also an effective tool for electricity distribution and reliability. Additionally, a microgrid has the capacity to disconnect from the main grid and be self-sufficient for a period of time but it can also remain connected and function alongside a larger grid system in normal operations. This is essential in a blackout or disaster scenario since a microgrid can disconnect from the other supply issues or even equipment damage that could be occurring elsewhere. This allows the microgrid to avoid cascading failures and provide reliable power in its specific service area [6].

The focus of this paper will be on the microgrid's ability to disconnect from the larger electric grid in a time of outages and be able to reliably provide power to a specific section otherwise referred as an island state. However, this requires that a microgrid have its own energy management system (EMS) and far more refined control methods than a traditional EMS since both the energy demand and consumption is at a far more granular level [7]-[10]. These enhanced requirements are implemented in this paper with two systems. Firstly, the day-ahead scheduling is used to optimize resource allocation since an emergency usually unfolds on a day-to-day basis. This system also makes sure that demand is being met. Lastly, it also allows cost approximation to allocate the correct energy supply ensuring effectiveness and ideal dispatching [11].

In addition to physical infrastructure, new forms of EMS including intermittent energy such as solar panels must be considered for resource allocation [11]. Microgrid functionality must be built into the system as more microgrids are being integrated or being developed alongside the main grid. This will have far reaching consequences in energy management systems as large changes in both the generation and consumption of energy are rapidly shifting.

The energy management system in a regular electrical system has incredible reliability and is a marvel of the modern world. Unfortunately, this reliability and interconnectedness is only guaranteed for normal conditions. The electric grid's ability to respond to issues under abnormal conditions such as storms, flooding, or other disasters may be reduced [6].

This paper primarily focus on such circumstances where the normal standards for reliability are not available. The high standard is only possible due to a vast and durable interconnected system which relies on large-scale generation transmitted to distributed residential systems. These infrastructure advantages are guaranteed in a natural disaster where due to damage, the system can be disconnected into multiple sections. When this happens, individual residential homes or industrial systems must have previously installed redundant energy resources such as diesel generation or BESS. Otherwise, their ability to receive electricity is entirely dependent on the speed at which the whole system can be reintegrated into a default state [12].

Therefore, advanced EMS software is necessary along with more resilient physical assets to harden the overall grid [1]. There are also new forms of distributed generation which change the dynamics of power transmission. All these factors require a rethinking of acceptable risk which currently is not acknowledged for existing systems. This paper utilizes day-ahead scheduling with specific time segments by assigning certain cost objectives to various resources including solar power, load curtailment, BESS, and diesel generation. This allows the model to create the most effective mix of resources to supply a load while minimizing resource usage throughout the day.

This paper presents one such approach to reduce unreliability by looking at day-ahead scheduling resource allocation which is then analyzed through a risk management method specifically a conditional value at risk (cVaR) analysis method to determine the risk factor of load curtailment throughout the day. This framework points out how intermittent resources and non-critical load curtailment can increase reliability [13]–[14]. The goal is to understand that not only load curtailment can be necessary in certain situations but how to quantify this necessity to ensure that system reliability is maximized in an emergency. It also creates a starting point to discuss instances where property that is currently controlled by individual use can be used in a more communal manner. This will allow a more sophisticated conversation about non-critical load curtailment instead of the current reality of demand reduction occurring haphazardly [15].

The remainder of the paper is organized as follows. Section II presents and describes the mathematic model for microgrid optimal scheduling. Section III presents the proposed cVaR analysis framework. Case study is presented in Section IV. Finally, Section V concludes the paper.

## II. MATHEMATICAL MODEL

The objective function in this paper is to maximize power availability for priority customers by minimizing risk and cost of volatile power generation sources.

$$\min \sum \{C_{B_{N_t}} P_{BESS_t} + C_{fuel} P_{Di_t} + C_{D_e} D_{e_{c_t}} + U_r C_{batt,red}\} \quad (1)$$

The objective function represented by (1) is a variation of the cost function of traditional unit commitment models showing resource allocation for BESS, diesel, and load curtailment while balancing demand and PV generation.

$$D_{Load_t} - D_{e_{c_t}} = P_{Di_t} + P_{BESS_t} + P_{PV_t} \quad (2)$$

Constraint (2) represents a basic requirement for all electric grid operations ensuring that demand meets supply. The usage of $D_{e_{c_t}}$ to minimize demand will be explained in the Load Curtailment section. Diesel systems are a useful fuel source around the world in grid operations as a DER alongside BESS and residential PV [3]. As a base constraint, there is a maximum discharge and charge capacity for diesel generators to meet technical limitations as

$$P_{Di_{min}} \leq P_{Di_t} \leq P_{Di_{max}} \quad (3)$$

Equation (4) defines the fundamental connection between how demand is configured in the system. Constraint (5) sets the grouping of priority customers and essential customers. Priority customers are a fraction defined by $\varepsilon$ of the essential customers. In the essential customer group, only $D_{e_{c_t}}$ is defined as essential customer curtailed are removed from the system as shown in (6). Equation (7) limits the $D_{e_t}$ and $D_{e_{c_t}}$. (8) – (10) enforces the BESS status to be charging, discharging or idle. Constraints (11)-(13) limit the charging and discharging power. Equation (14) defines the cost factor for any usage of the battery outside the green zone.

$$D_{Net\ Load_t} = D_{p_t} + D_{e_t} - P_{PV_t} \quad (4)$$
$$D_{p_t} = \varepsilon D_{e_t} \quad (5)$$
$$D_{p_t} + D_{e_t} - D_{e_{C_t}} = P_{Total_t} \quad (6)$$
$$D_{e_t} \geq D_{e_{C_t}} \geq 0 \quad (7)$$
$$U_{C_t}\{1, charging\ state.\ 0, not\ charging\} \quad (8)$$
$$U_{D_t}\{1, discharging\ state\ 0, not\ discharging\} \quad (9)$$
$$U_{C_t} + U_{D_t} \leq 1 \quad (10)$$
$$P_{BESS_t} = P_{D_t}^B - P_{C_t}^B \quad (11)$$
$$0 \leq P_{C_t}^B \leq U_{C_t} P_{C_{max}}^B \quad (12)$$
$$0 \leq P_{D_t}^B \leq U_{D_t} P_{D_{max}}^B \quad (13)$$
$$\begin{cases} SOC_{min}^{green} \leq SOC_t \leq SOC_{max}^{green}, & U_r = 0 \\ SOC_{min}^{green} > SOC_t\ or\ SOC_t > SOC_{max}^{green}, & U_r = 1 \end{cases} \quad (14)$$

The cost of the battery system is connected to the maximum life cycle to calculate the overall cost of the battery as connected to cycle count. This allows us to take a specific portion of battery usage such as one day and connect it to the overall cost of the battery by (15). $N_{bat_t}$ in (16) is the number of cycles the battery is at while $\lambda_{N_{bat_t}}$ is the capacity factor loss at the Nth cycle. Equation (17) defines the total cost of the BESS and (18) represents the difference of the degradation cost between different time intervals.

$$N_{bat_t} = \sum_{t=0}^{t} \frac{1}{2}(DoD_t + DoC_t) \quad (15)$$
$$N_{bat,max_t} - N_{bat_t} = \frac{N_{bat,max_t}(1 - \lambda_{N_{bat_t}})}{\lambda_{N_{bat_t}} * \gamma} \quad (16)$$
$$C_{B_{N_t}} = \lambda_{N_{bat_t}} * C_{B_{Total}} \quad (17)$$
$$\Delta\lambda = \lambda_{N_{bat_t}} - \lambda_{N_{bat_{t-1}}} \quad (18)$$

III. PROPOSED CVAR FRAMEWORK

This section explains how the costs defined in the model for day-ahead scheduling is used in the cVaR framework. Fig. 1 presents the process from day-ahead scheduling to cVaR analysis. First, all the scenarios in the day-ahead scheduling must be completed. This means that for one time interval, t, there will be hundreds of scenarios operating with different demand constraints and PV generation. Then, when all N scenarios have been completed, they will create a large set of data points of cost optimized resource allocation including any possible load curtailment. These load curtailment measurements can be tested for stability and resiliency and used to create a risk profile using cVaR analysis.

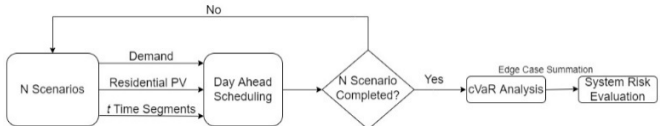

Figure 1. Procedure of the proposed microgrid scheduling and risk analysis.

*A. cVaR Formulation*

The use of risk-constrained scenarios in financial models and utilities is to maximize profit with an internal pricing mechanism [16]. cVaR is a popular risk calculation algorithm. It is built on the work of value at risk (VaR) which calculates how to reduce risk within a certain confidence level (β) by minimizing loss due to the uncertainty in specific variables [16] otherwise defined as equation (19). The $f(x,y)$ factor in (19) is defined as the losses calculated. The x denotes the variables available to fine tune and reduce risk where y represents the volatile uncertainty inherent in our system. By minimizing the worst-case scenario of y, the system could create an expected risk profile. This is calculated by taking the smallest possible cost (α) that is greater or equal to $f(x,y)$ and then calculated the risk factor over β. This can be used to calculate the level of risk inherent in investing in certain markets and diversification tools (such as cash or bond hedging).

$$VaR = min\{\alpha \in R: P\{f(x,y) \leq \alpha\} \geq \beta\} \quad for\ 0 \leq \beta \leq 1 \quad (19)$$

Unfortunately, VaR suffers from two key issues. Mathematically, it has a lack of convexity and subadditivity making it non-ideal for intensive calculation operations. Secondly, VaR only minimizes losses within a given confidence level and does not consider losses occurring at a confidence level outside of its boundaries at 1-β. cVaR allows a better grasp for situations where a small likelihood of risk could have a huge effect [17]. cVaR as a financial constraint is seen in equation (20).

$$cVaR = \mathbb{E}_y(f(x,y)|f(x,y) \geq VaR) \quad (20)$$

In this evolution of the original VaR equation, the cVaR is now taking the expected value of random variables above its VaR consideration. In other words, it is taking the loss factors inherent in the system and calculating them in situations of 1-β or above the standard confidence interval. This is a much more robust and flexible system since it allows forecasting of situations where non-likely events outside of the confidence interval occur. Additionally, a higher cVaR means the system is inherently less stable because in non-normal situations, the losses can be considerably higher. To transition from the above equations to models with samples, (20) can be converted into equation (21).

$$cVaR = min\left(\alpha + \frac{1}{N(1-\beta)}\sum_{i=1}^{N}[f(x,y) - \alpha]^+\right) \quad (21)$$

The first shift here is the addition of N moving the model from continuous to samples with N scenarios. The second change is that the positive component of our losses taken by known x and volatile y subtracted by α as our hedging cost. For our formulation, we can then replace $[f(x,y) - \alpha]^+$ with $z_i$ as shown in (22). The cVaR equation can now be redefined with $z_i$ as seen in (23).

$$z_i = [f(x,y) - \alpha]^+ \quad (22)$$
$$VaR = min\left(\alpha + \frac{1}{N(1-\beta)}\sum_{i=1}^{N} z_i\right) \quad (23)$$

*B. cVaR Application in Microgrid*

This section explains how cVaR will be used to maximize power reliability for priority customers. cVaR gives a weighted average of risk above the normal confidence level. This allows a calculation of the risk in high-demand scenarios that can occur in emergency situations. Equation (24) takes $f(x,y)$ from (19) and defines the combined losses as demand subtracted from diesel, PV, and BESS [14]. $\alpha$ in (25) is set as the smallest load curtailment while maintaining stability at the confidence level $\beta$ [16]. The $\alpha$ value will be measured in units of kilowatts. While keeping with cVaR convention, demand load that is curtailed will be referred to as $\alpha$ moving forward. All this can be represented as:

$$f(x,y) = D_{Net\ Load_t} - P_{BESS_t} - P_{Di_t} \quad (24)$$
$$z_i = [D_{Net\ Load_t} - P_{BESS_t} - P_{Di_t} - \alpha]^+ \quad (25)$$

## IV. CASE STUDIES

The test residential microgrid is designed with currently available commercial products. It is designed with a battery system made up of twenty Tesla Powerwall batteries with a capacity of 15 kWh each that starts at an initial value of 10 kWh for each battery. $B_i$ is the capital cost of the BESS system at $10,000 [18]. The standard rooftop residential solar output is at 4 kW during peak solar generation [19]. There are ten residential homes in need of power all with installed solar panels. $\varepsilon$ in (5) is set to 0.5 therefore priority customers were a total of 33% of total demand. This means that a maximum of 66% of customers can be essential customers [11]. A simulation of all 187 possible scenarios, $N$, is run and the battery, diesel, and PV combination are recorded for each specific segment. $\beta$ is defined as 5% for the confidence level in this analysis.

The input data for the scenarios including load demand and PV generation is graciously provided by Pecan Street. This is part of Pecan Street's Dataport Project [19] which includes the world's largest resource for residential energy use data, electric transportation and has been expanded to include residential water usage, electric transportation, and regenerative agriculture [20]. Electricity demand as well as PV generation will have expected statistical deviation from historical data. $SOC_{min}^{green}$ is defined as 20% and $SOC_{max}^{green}$ as 80% in the model using values from previous research [21]. $P_{Di_{min}}$ is set as 0 and $P_{Di_{max}}$ is set as 3.75 kW for the system generator. The diesel generator is assumed to have sufficient fuel to operate during the whole course of the day. $P_{D\ max}^B$ and $P_{c\ max}^B$ is defined as 5 kW for one Tesla Powerwall [18]. $\Delta T$ represents the length of time segment which is 15 minutes in this paper. There are 96 segments for a 24-hour period. The load curtailment if any for each fifteen-minute interval is recorded. The load curtailment is divided by the total demand supplied and recorded in a matrix. Python is used to take these values and calculate the conditional value at risk for the most demanding and highest load curtailment of the five percent of scenarios (nine scenarios) of the total set of 187 scenarios for all time segments.

The results highlight the cVaR analysis on the microgrid system for one full day or 96 segments on a total of 187 scenarios. Fig. 2 shows in how many instances curtailment was necessary in the model. This showcases a high level of self-sufficient reliability that above would be a boon to the existing electrical grid infrastructure. The system had zero instances of load curtailment for 90% of scenarios. It had a maximum of 13 instances of load curtailment in the most challenging 5% of cases for all segments.

From a system wide load curtailment view, now we can take a more in depth look at the 5% of challenging scenarios in terms of balancing generation and demand. The standard deviation shown in Fig. 3 presents the difference in values of the dataset for each segment.

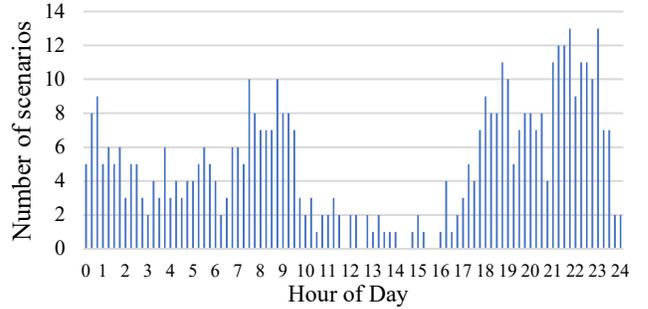

Figure 2. Time segments with Active Curtailment.

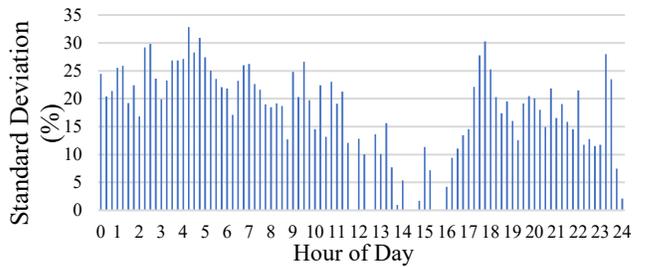

Figure 3. Standard deviation of load curtailment in cVaR analysis.

Within each segment, there is a 20-30% standard deviation indicating the model is robust. These results prove that the model can take in very different demand constraints and respond appropriately to the need of the specific scenario. Interestingly, the standard deviation is largely consistent throughout the day, indicating that the load curtailment deviation is not too different between sample segments. An exception to this is late mornings to end of the afternoon when the generation of residential PV are sufficient, there are far less load curtailments and therefore the standard deviation is lower.

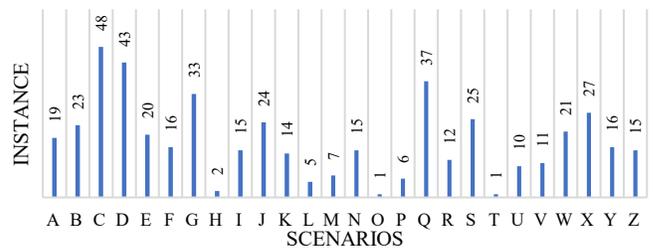

Figure 4. Active curtailment during an entire day.

Fig. 4 presents the twenty-six scenarios or 13.9% of the entire scenario dataset that was responsible for all load curtailment. This is expected since the model was tested on a robust dataset which has microgrid scenarios with larger than expected demands. This is very likely in emergency situations due to weather conditions, and it is important to note how the microgrid would react in these scenarios.

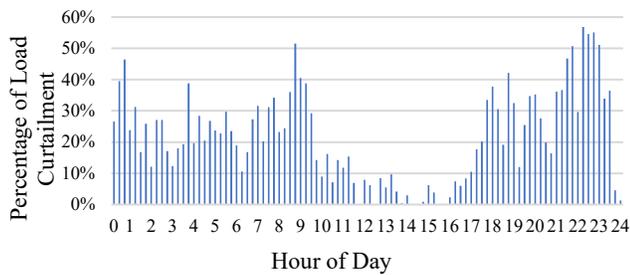

Figure 5. The cVAR analysis at a 5% confidence level.

As shown in Fig. 5, the behavior of the case study matched expectations in the following ways. The risk when calculating real time energy management for the hours of 10 AM to 4 PM were reduced and in some time-segments brought to zero. This means residential solar times matched demand at these times and reduced risk of load curtailment. This is one of the main benefits of residential solar. It especially helps microgrids in providing a power source for a part of the day. Load curtailment was expected to be used in a small percentage of the case. The correct and targeted load curtailments can improve the system's reliability for priority customers. This is complementary to grid hardening efforts but has the advantage of lower costs because it can be built with existing infrastructure.

V. CONCLUSIONS

cVaR analysis is conducted in a stand-alone microgrid alongside day ahead scheduling in this paper. The proposed energy management system demonstrates the adaptability of a multitude of generation sources being utilized along with load curtailment in different demand-constraint scenarios. The objective was to conduct a risk assessment on a microgrid system to assess likelihood of load curtailment. This allows for evaluating the risk of existing system infrastructure facing controlled load curtailment in a disaster scenario. Instead of proposing a brand new microgrid installation, existing electrical infrastructure in neighborhoods particularly those with high residential penetration can be retrofitted with additional diesel generation and battery storage services alongside its own energy management system with the proposed energy management system.